# Mobile Recommender Systems: An Overview of Technologies and Challenges

Nikolaos Polatidis
Department of Applied Informatics
University of Macedonia
Thessaloniki, Greece
polatidisn@acm.org

Christos Georgiadis
Department of Applied Informatics
University of Macedonia
Thessaloniki, Greece
gxri@acm.org

*Abstract*— **The use of mobile devices in combination with the rapid growth of the internet has generated an information overload problem. Recommender systems is a necessity to decide which of the data are relevant to the user. However in mobile devices there are different factors who are crucial to information retrieval, such as the location, the screen size and the processor speed. This paper gives an overview of the technologies related to mobile recommender systems and a more detailed description of the challenged faced.**

## I. Introduction

The evolution of computers in combination with the rapid development of mobile phone platforms such as the Google Android and the Apple iPhone in conjunction with the need for m-commerce need has brought up the use of personalization and recommendation systems. The use of the internet is moving forward with a direction towards mobile devices and although such devices come in various forms, in this paper we refer to smartphones that can access the internet and have features to enable access to information through a suitable way. Moreover, mobile devices have limitations related to hardware capabilities, network connections and usability issues such as the screen size [15].

Mobile devices have become very popular nowadays and they are considered to be a primary access information environment. However as the information on the internet grows and the people who use these devices become larger and larger there is a need to face the challenges that are tight related to mobile computing environments. The need to face the information overload is the most important nowadays in combination with the limitations that most of these devices have (such as the size of the screen, the processor speed and memory capabilities), direct us to the use of recommendation technologies.

Recommender systems are concerned with the dynamic customization of data received over the World Wide Web and are based to user preferences. The scope of the recommendations is to assist the user to decide what to buy, who to make friend to a social network or what news to read. Due to information overload on the internet, personalization systems are one of the most valuable tools [12]. Additionally it should be noted that it is a very demanding process to design and develop such a system, since it combines knowledge and skills from different computer science fields. Despite of that, a number of well-respected methods have been developed the past few years, with some of them being used in commercial environments. Moreover, in mobile devices the information access problem becomes even harder because of the difficulties found.

It is important to note that the algorithms applied to personal computers cannot be transferred directly to a mobile device, since there are different needs, special characteristics and limitations. The needs are about m-commerce, location-based services found mainly in tourism and mobile financial services. Characteristics refer to the user interface, processor, memory capabilities and limitations are about the network boundaries found in GSM, Wi-Fi and GPS systems. However the advantages are more important. These include the ubiquity and the location-based service. These are the crucial factors that mobile recommender systems are based on most of the time.

Furthermore the need for privacy has become a very important aspect of mobile phones that use personalization techniques. It is vital for the system to use some private data in order to provide accurate recommendations. However it should be taken into consideration that privacy is a massive problem with negativity towards the use of recommenders in mobile environments [4, 9, 10]. Most of the time simple users are not aware how e-commerce organizations use these data and they react in various destructive ways. We have reached a point that merchants want to improve their service and use unfair practices. However, there is a reconciliation point that could be reached if both parties are willing to work towards this road.

The research is focused on how the users can receive personalized content in their smartphones using current technologies. However, in mobile environments attributes such as location and time should be embedded to such algorithms but on the other hand there are privacy concerns that have to be taken into consideration. Moreover, a mixture of current recommendation technologies that study the special mobile characteristics and use data from social networking websites will be discussed, along with technologies that can be used to solve the privacy problem.

Additionally, there are still open questions that need extensive research to be answered. These include the goals of mobile recommender systems and the expectations of the users, if there are any type of implications associated with the location-based parameter and if this should be compulsory or enforced only when questioned. Also, it is not clear if there is a mobile domain or there is a number of e-commerce scenarios that is more suitable to mobile devices such as tourism. However these questions are still open, because mobile computing is a relatively new field of study in computer science and the lack of surveys and results is obvious [11, 15].

In addition to the above, an issue in mobile and ubiquitous computing is context awareness. By context is denoted the environment parameters and if they can play a role in recommendation results. The context usually adds additional parameters to the running algorithm which include demographics, objects, surroundings or other factors that could be considered relevant by the user such as health, mood, social activities and application context like web browser history and email mining. Also into consideration the sequence of visits can be taken in order to find a pattern and make predictions [7, 15, 24]. The application areas that widely use context awareness at the moment include, but are not limited to, m-commerce and tourism.

## II. RELATED TECHNOLOGIES

Mobile recommender systems are usually found in electronic commerce environments. These environments have changed rapidly within the last few years and the growth of the internet, the wireless networks and the development of mobile devices and environments has brought new terms in our lives. E-commerce, M-commerce and recommender systems are some of these new terms and are introduced below.

### A. E-commerce

E-commerce or else known as electronic commerce, is the new way of doing business over the internet. It is concerned with the transactions made over the internet and involves different types of business conducted such as buying, selling, online banking, bill payments, job seeking and offering, new reading, social networking, travel services, auctions and real estate among many others. E-commerce has many benefits to both the organizations and customers. The most important benefits include making business at any time and from any possible place and financial gaining for both the business, which reduces the cost and the customer who buys cheaper. E-commerce systems employ recommender systems widely to improve sales [7, 16].

### B. M-commerce

M-commerce or else known as mobile commerce, is a new way of e-commerce done in a mobile device using a wireless network. A common example is using a smartphone or a tablet to access the internet over a Wi-Fi network. Types of commerce over mobile devices include mobile banking, content purchase, news reading, auctions and location-based services mainly used for tourism purposes [7, 16]. M-commerce technologies widely use recommender systems as well, but customized to its needs.

### C. Recommender Systems

Recommender systems are computer algorithms used widely in e-commerce to suggest items to a user. The recommendations are about what items to buy, news reading, social networking connections and what movies to rent among many others. Among the most popular websites that use recommender systems is Amazon.com, which provides a personalized web page to each individual user. Netflix is another example website that uses recommender systems to suggest movies and TV shows. Such systems in general suggest a list with N top items relevant to the user. The items are retrieved according to rules set by the algorithm and suggest the topmost from the list, depending on the interface. Recommender systems were developed to make daily decisions simpler. These decisions are mostly about low cost environments such as book and movie suggestions, with their primary scope being to relief the user from long searches [11, 12, 15].

Recommender systems is a relatively new field of study in the computer science and e-commerce literature, however its techniques are widely adopted and solved well to a level the information overload problem. Moreover nowadays with the rapid growth of mobile devices and their operating environments in conjunction with the evolution of mobile networks the presence of the internet is everywhere, thus making the use of recommender systems in mobile environments a necessity [5, 7].

Furthermore, the use of recommender systems is essential for the service providers and not only the users. The reasons that the providers use such systems are:

#### i. Increase Sales

The most important reason for a commerce site and mobile application to use a recommendation technology is to increase its sales and revenue. This is accomplished because the recommender usually suggest items that are relevant to the user, according to his history and preferences.

#### ii. Employ diversity of items

A recommender system would usually suggest items from a large range that otherwise the user would be very difficult to find. Therefore the algorithm will suggest different items, increase the sales of items that otherwise is challenging to sell and increase the total sales and profit.

#### iii. Increase user satisfaction

The user is more satisfied by the overall service offered and it is likely to suggest it to others.

#### iv. Increase loyalty

It is more likely for a user to revisit a site or reuse a mobile application if he is satisfied with the quality and treat.

As a final point, it is noted that a recommender system despite the algorithm that uses, it takes as input personal information from the user it creates the recommendations either locally or in a distributed environment and it passes the predictions to the interface of the services the user is using. Furthermore the three most important and widely used recommendation algorithms are the following.

*1. Collaborative filtering*

In collaborative systems the basic idea is to find which users share the same interests with you in the past. The main idea of these systems is that the users who had a taste similar to you, likely will have the same taste in the future. Pure collaborative techniques take a user-input matrix with ratings as the only input and generate a prediction value indicating similarities to other users [**11, 12, 15**].

The following table represents a ratings database for Alice and four other users.

TABLE 1 A ratings database

| Users | Item1 | Item2 | Item3 | Item4 | Item5 |
|---|---|---|---|---|---|
| Alice | 5 | 3 | 4 | 4 | Null |
| User1 | 3 | 1 | 2 | 3 | 3 |
| User2 | 4 | 3 | 4 | 3 | 5 |
| User3 | 3 | 3 | 1 | 5 | 4 |
| User4 | 1 | 5 | 5 | 2 | 1 |

A number of similarity methods exist such as the cosine-based similarity and the adjusted cosine similarity. Also the Pearson correlation based similarity has been widely used. Each method returns a number from-to, like-dislike value [**15**].

*2. Content based filtering*

Content based recommenders are somewhat simpler to implement, since they are based on metadata of the actual data. This metadata can be some technical description of an item, the genre of a movie, the title, type, author or other defined set of keywords [**11, 12, 15**].

This is done by maintaining a list of attributes and searching within the list. Content based recommendations is a durable technique and the database is constantly updated with user preferences. Moreover content based is used in systems like news filtering and information retrieval. Furthermore systems like that, are used when there are not enough ratings to perform a collaborative based approach.

*3. Knowledge based filtering*

In practice most systems are based in collaborative filtering methods, a technique which is based on other user ratings only [**11, 12, 15**]. On the other hand, content based approaches use metadata information, such as movie categories or keywords. The main benefit of these methods is the low cost to acquire the data from the users. However there are many circumstances that these methods are not sufficient. Collaborative and content based methods usually perform well in low cost environments such as books and movies, where other users provide ratings and information more often [**11, 12, 15**]. In other situations such as buying a personal computer or a digital camera a knowledge based algorithm shall be used.

As it is distinct by its name, knowledge based systems rely on data provided by the user and use them as constraints in order to provide recommendations.

TABLE 2 Example digital camera characteristics

| Name | Price | Mega Pixels | Zoom | Screen size | Quality |
|---|---|---|---|---|---|
| Camera1 | 100 | 6 | 2x | 2cm | Low |
| Camera2 | 119 | 8 | 2x | 2.5cm | Medium |
| Camera3 | 200 | 12 | 4x | 3cm | High |
| Camera4 | 150 | 10 | 3x | 3cm | Medium |
| Camera5 | 140 | 8 | 4x | 2.7cm | Medium |

The table above describes the characteristics of a dataset about digital cameras as found in the database of a system. The user can then add specific rules to the system in order to receive personalized recommendations.

Consider the following example described using logic rules:

*Price <=150^MegaPixels>=10^Quality>=Medium*

The system will list all cameras with their price being lesser than or equal to 150 and their mega pixels greater than or equal to 10 and their quality being at least medium. In this case camera4 will be listed since it is the only one satisfying the conditions.

*4. Hybrid Algorithms*

Hybridization is the combination, or the use of all, features from different algorithms in order to make a recommendation. While each of the algorithms described above is dominant in their respective field, the need for hybridization is gaining more attention [**2, 3, 15, 25**]. Moreover hybrid algorithms have been categorized according to their design [**15**].

*1. Monolithic*

A monolithic algorithm integrates different features from each algorithm. It could include every feature that an algorithm supports or a number of them.

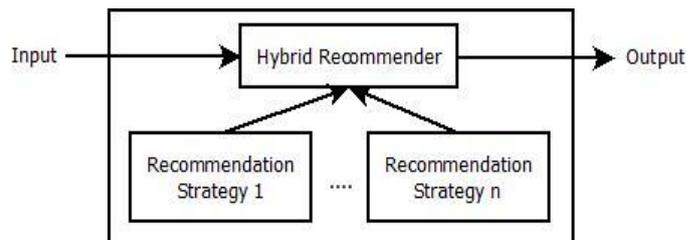

**Figure 1.** Monolithic Recommender Approach (Adapted from Recommender systems, An Introduction)

*2. Parallelized*

A parallel algorithm runs each selected algorithm in parallel, takes the output from each one and passes it to a predefined hybridization stage.

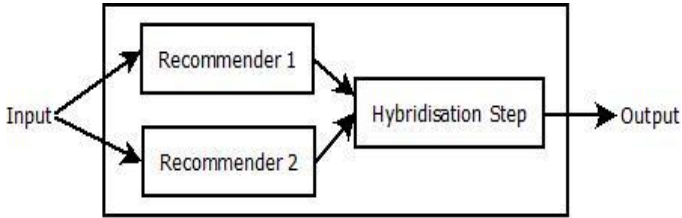

**Figure 2.** Parallel Recommender Approach (Adapted from Recommender systems, An Introduction)

*3. Pipelined*

A pipelined algorithm runs each algorithm in sequence and takes the output from the previous one and uses it as the input for the next one until the required outcome has been achieved.

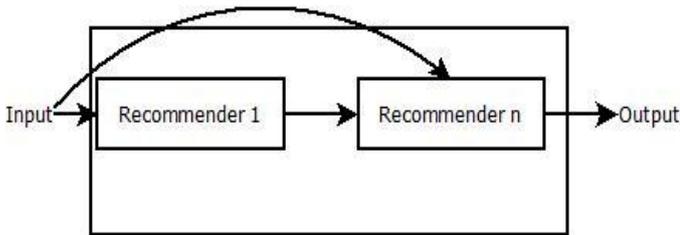

**Figure 3.** Pipelined Recommender Approach (Adapted from Recommender systems, An Introduction)

### III. MOBILE ENVIRONMENTS

In the internet era and with personal computers being everywhere the need for e-commerce and recommender technologies is essential. Likewise, smartphones and tablets use the same technologies in their respective environments with extra parameters taken into consideration. Such parameters include location, time and screen limitations. The information overload is so high nowadays, that recommender systems are necessary in almost every aspect of mobile devices. Furthermore, examples of recommendations include mobile commerce, news reading and finding related services (such as hotels or other tourism related information). Additionally, the personalization of the operating system itself has been taken into consideration [**5, 7**].

Moreover mobile devices have progressed so fast that tend to become the primary source of access to social networks [**5**]. Users who want instant access, from everywhere and not use a computer to do that, tend to use a mobile device such as a smartphone or tablet [**5**]. The network communication facilities such as the cellular, Wi-Fi and GPS have aided towards this direction.

The progress both in hardware and software have enabled mobile devices to offer extraordinary services and help people to achieve experiences that otherwise would be unknown to them. Context-awareness is the most important aspect that drives mobile devices [**1, 5**]. Mobile devices are based to services in order to support their users. These services are location-based, user-based (such as the use of personal profiles), device-based (such as the memory, processor speed and screen size), spatio-based attributes (such as day, night, mood and/or weather conditions) and social-based attributes based on communicating with other users [**15**].

In mobile devices there are other factors than the traditional, which should be taken seriously in order to improve recommendations. Such elements include access to the e-mail agent of the OS, to the call logs, SMS excavating, internet and online chat services history and the use of a user well-defined profile [**6, 8**]. Additionally, a very serious aspect is the use of data from social networks, which nowadays its use is put forward mobile operating systems that have integrated features of networks such as Facebook and twitter [**1, 17, 20, 23**]. Also, the use purchase history data from large auction sites such as EBay must be taken into consideration.

A very important aspect in respect to the usage of mobile devices and recommendation system to be more specific is privacy. It is the main cause that mobile users avoid to use such technologies. It is a challenge to the use of recommenders in mobile environments and it is discussed further in the challenges section below.

### IV. CHALLENGES

Below we discuss challenging topics found in mobile recommender systems literature. These areas are still under active research and serious future work is required.

*1. Privacy*

Privacy is about ensuring that the user data will be kept private whatsoever. Privacy policies in conjunction with the knowledge level of the users about the subject guide them towards a negative behavior when they are being asked about to pass data in order to receive more personalized content [**4, 10**].

In recommender systems the users are divided in three main categories, with regard to their judgments and choices [**10**].

i. *Users that will give any kind of information to a recommender system with return more personalized content.*

ii. *Users that will give some information to a recommender system in order to get improved recommendations.*

iii. *Users that will not give any kind of information to a recommender system because of privacy concerns.*

The category that passes some information include general data such as gender, age, education. These data are given easier rather than more specific personal data [**10**]. However to improve recommendations we have to convince the users that their data will be safe and conduct research about privacy.

This research can be done either in a lab using observation techniques, either directly or indirectly and by asking the users to answer questions about. Moreover the benefits of personalization should be made very clear to the user and provide them with a very clear, certified, privacy statement and seal. The users want to know how their data will be used.

Additionally, other less important factors that play a vital role is the confidence towards a website and app in the case of a mobile device, any positive previous experience from a vendor, the design and reputation of a particular website or mobile app. Last it should be noted that earlier research has shown that the average user would disclose information very easily in exchange for money or gifts [10]. Some technical solutions that could partially solve the problem include the use of pseudonymous users and profiles, including subcategories, client-side personalization and distribution techniques [6]. Finally the use of an external privacy tool in the form of an app can be used to control privacy.

### 2. Mobile App Permissions

Mobile applications and their permissions is critical part of privacy as well. Apps have the ability to access and share user data and possibly amend them. Moreover, apps have the ability to communicate with each other, thus raising privacy and security concerns. Conventional privacy methods are not valid here and other aspects must be taken into consideration. To make it more clear, an app can use the permissions granted to another app in order to gain access to data that otherwise it wouldn't be able to. However, app with the same UID only can perform operations on each other's permissions [9].

### 3. Social Media Integration

The integration of social networks in mobile operating systems in conjunction with the growth and speed of internet has brought huge amounts of social networking data. It is one of the easiest ways to collect data from users that they are willingly disclose and can be very beneficial to e-commerce and businesses in general. However it is a challenging procedure to collect and separate the necessary data used for e-commerce purposes. Data from social networks shall be used to assist customers by making more accurate recommendations. Social media allow the exchange of information in different categories, each one having special characteristics. The most important categories are the following [18]:

i. *Social Networks*. These are web-based networks of users that connect and interact with each other through updates, online chat and multimedia sharing. Well-known examples include Facebook and twitter.

ii. *Blogs*. These are websites that contain text and multimedia content that is arranged in a chronological order. These are usually maintained by a single user or a team of users. Famous examples include Blogger and Word Press.

iii. *News reading*. Selection of articles or categories to read from online specified news websites.

iv. *Online video sharing*. This includes the sharing of videos to services such as YouTube or Veoh.

v. *Online photo sharing*. This is photo sharing in services such as Flickr or Picasa.

vi. *Auction sites*. Data from history purchases and search history at sites such as EBay.

A huge amount of data is created on each of these social media networks daily and this is a trend that is growing exponentially. As an indicative figure, let us mention that the number of Facebook and Twitter users increased by 112% and 347% respectively from January 2009 to January 2010 [1]. However, it is a challenge to decide which data to use. Moreover there is a number of spammers that create more data than real users [1].

All social networks have application programming interfaces (APIs) that can be used to communicate with them. However it is a challenge to retrieve the data and respect privacy as well. Apart from information related to the user such as gender, age, background, relationship status, there is a number of dynamic contents that is constantly changing. These contents include mood, location, posts and posts made by other users on his personal page [22].

### 4. Enhanced Role-Based Access Control Model

An edge point that is directly related with privacy, which at the moment is the highest challenge in mobile environments, is Role-Based Access Control (RBAC). A role in an organization is a set of rules describing what actions a particular person can take within. Therefore it is clear that a novel role model is necessary to determine the access level and constraints. Even though a satisfying number of technologies exist, due to an always moving forward environment there are always new challenges [13].

Role-Based Access Control is the fundamental security model enforced nowadays. It is used in various ways to define the ways that users will have access to resources. Moreover, with the right modelling can be used to control app permissions as well.

The two most important aspects of access control are authorization and authentication [13]. Authorization is if the user has the permission to perform certain operation and authentication is to identify if the user is the one who claims he is.

An abundant challenge is the proposal of new modified RBAC model, which is essential at this point. This new model should be able to maximize the level of privacy of a mobile recommender system. The difficult part is to define the constraints that the new model should cover in terms of users, operations and permissions of both users and applications. Other important characteristics could be the least privilege technique and group access control relationships. In least privilege, the user should be able to perform only the operations required and not any other either at an above or at below level whatsoever. In group access, a user of a group could have additional permissions that other users in the same group do not. Furthermore a number of crucial operations could be performed only if two users at the same time are logged in to perform the operation.

## V. FUTURE WORK

Current models provide practically no protection for privacy and data security with a clear example being Facebook that requires to associate users' personal information with their id [19, 21]. Recommender systems in mobile devices should use a trusted centralized server in order to achieve privacy. This can be succeeded with the use of a modified Role-Based Access Control model and the trusted platform module. The RBAC model needs to be extended to provide enhanced security both in the permissions of the mobile applications and the security of the data itself in the server.

Cloud computing is another feature that should be taken into consideration. Cloud computing platforms can be used to store user profile information. This information can then be joined with the mobile device to improve the user experience and provide a low-cost environment. Moreover, this carries privacy and security concerns that can be resolved with the use of the extended RBAC model.

## VI. CONCLUSIONS

Although recommender systems are in every personal computer and mobile device nowadays, there is a number of factors that users of mobile platforms in particular take into consideration and avoid their use. These factors are tight integrated to privacy. Additionally, a new trend in the internet era is social networks and its derivatives with huge amounts of data exchanged every day. These data should be used in recommender systems to improve personalization. As a final important point is the development of new Role-Based Access Control (RBAC) model that will improve privacy ambiguities found in app permissions and user access to data.

## REFERENCES

[1] N. Jabeur, S. Zeadally and B. Sayed. "Mobile Social Networking Applications". Communications of the ACM, Vol. 56, issue 3, pp. 71-79, 2013.
[2] M. A., Ghazanfar and A. Prugel-Bennett, "A scalable, accurate hybrid recommender system". Proceedings of the 2010 Third international conference on Knowledge Discovery and Data Mining, pp. 94-98, 2010.
[3] A. Gunawardana and C Meek, "A unified approach to building hybrid recommender systems". Proceedings of the third ACM conference on Recommender Systems, pp. 117-124, 2009.
[4] S. Lam, D. Frankowski and J. Riedl. "Do you trust your recommendations? An exploration of security and privacy issues in recommender systems". Proceedings of the 2006 International conference on Emerging Trends in Information and Communication Security, pp. 14-29, 2006.
[5] A. Oulasvirta, T. Ratternbury and E. Raita, "Habits make smartphone use more pervasive". Personal and Ubiquitous Computing vol. 16, issue 1, pp. 105-114, January 2012.
[6] W. Paireekreng and K. Wong, "Mobile Content Personalization Using Intelligent User Profile Approach". Proceedings of Third International Conference on Knowledge Discovery and Data Mining, pp. 241-244, 2010.
[7] F. Ricci, "Mobile Recommender Systems". International Journal of Information Technology and Tourism vol. 12, issue 3, pp. 205-231, 2011.
[8] D. Davidson and B. Livshits, "MoRePriv: Mobile OS Support for Application Personalization and Privacy". TechReport, Microsoft Research, Redmond, WA 98052, United States, 2012.
[9] B. Guillaume, B. Arosha, Y. Yijun, C. Jean-Noel and N. Bashar, "PrimAndroid: Privacy Policy Modelling and Analysis for Android Applications". IEEE International Symposium on policies for Distributed Systems and Networks, 2011.
[10] A. Jeckmans, M. Beye, Z. Erkin, P. Hartel, R. Lagendjik and Q. Tang, "Privacy in Recommender Systems". Social Media Retrieval Journal vol. 11, pp. 263-281, 2013.
[11] J. Konstan and J. Riedl, "Recommender Systems: from algorithms to user experience". Journal of User Modeling and User-Adapted Interaction vol. 22, issue 1-2, pp. 101-123, 2012.
[12] , F. Ricci, L. Rokach, B. Shapira and P. Kantor, "Recommender systems handbook", Springer, 2011
[13] D. Ferajollo, D. Kuhn and R. Chandramouli, "Role-Based Access Control", Artech House, 2nd edition, 2007
[14] W. Paireekreng and W. Wong, "The empirical study of the factors relating to mobile content personalization". International Journal of Computer Science and System Analysis vol. 2, issue 2, pp. 173-178, 2008
[15] D. Jannach, M. Zanker and G. Friedrich "Recommender Systems: An Introduction", Cambridge University Press, 2010
[16] K. Laudon and C. Traver, E-commerce, Pearson, 2013
[17] Y. Zhang and Y. Tong, "Mining Trust Relationships from Online Social Networks", Journal of Computer Science and Technology, vol. 27, issue 3 pp. 492-505, 2012.
[18] P. Gundecha and H. Liu, "Mining Social Media: A Brief Introduction." Journal of Operations Research, vol.9 pp. 1-17, 2012.
[19] L. Fang and K. LeFevre, "Privacy Wizards for social networking sites" Proceedings of the 19th International Conference on World Wide Web. ACM, New-York, pp. 351-360, 2010.
[20] A. M. Kaplan and M. Haenlein, "Users of the world, unite! The challenges and opportunities of social media" Business horizons journal, Vol. 53 issue 1, pp. 59-68, 2010.
[21] B. Krishnamurthy and C. E. Wills, "On the leakage of personally identifiable information via online social networks" ACM Computer Communication Review Vol. 40 issue 1, pp. 112-117, 2010
[22] H. Liu and P. Maes, "InterestMap: Harvesting social network profiles for recommendations" Workshop: Beyond personalization, San Diego 2005
[23] H. Ma, D. Zhou, C. Liu, M. Lyu and I. King "Recommender systems with social regularization" Proceedings of the fourth ACM International Conference on Web Search and Data Mining. ACM, New-York, pp. 287-296, 2005
[24] C. Palmisano, A. Tuzhilin and M. Gorgoglione "Using context to improve predictive modeling of customers in personalization applications". Knowledge and Data Engineering, IEEE Transactions, Vol. 20, issue 11, pp. 1535-1549, 2008
[25] M. Zhang "Enchasing diversity in top-n recommendations" RecSys '09: Proceedings of the third ACM Conference on Recommender Systems, ACM, New York, pp. 397-400, 2009.